\documentclass{osa-article}

\journal{oe}

\articletype{Research Article}

\usepackage{lineno}
\usepackage[T1]{fontenc}
\usepackage[utf8x]{inputenc}
\usepackage{hhline}
\usepackage{subfig}
\usepackage{float}

\begin{document}

\title{Measurement of the quantum-confined Stark effect in InAs/In(Ga)As quantum dots with p-doped quantum dot barriers}

\author{Joe Mahoney,\authormark{1} Mingchu Tang,\authormark{2}  Huiyun Liu,\authormark{2}  and Nicol\'as Abad\'ia\authormark{1,3,*}}

\address{\authormark{1} School of Physic and Astronomy, Cardiff University, Cardiff CF24 3AA, UK\\
\authormark{2} Department of Electrical and Electronic Engineering, University College London, London WC1E 7JE, UK\\
\authormark{3} Institue for Compound Semiconductors, Cardiff University, Cardiff CF24 3AA, UK}
\email{\authormark{*} abadian@cardiff.ac.uk}
\homepage{\authormark{*} https://orcid.org/0000-0002-7355-4245}


\begin{abstract}
The quantum-confined Stark effect in $InAs/In(Ga)As$ quantum dots ($QDs$) using non-intentionally doped and p-doped $QD$ barriers was investigated to compare their performance for use in optical modulators. The measurements indicate that the doped $QD$ barriers lead to a better Figure of Merit ($FoM$), defined as the ratio of the change in absorption $\Delta\alpha$ for a reverse bias voltage swing to the loss at $1 \: V$ $\alpha(1 \: V)$, $FoM=\Delta\alpha/\alpha(1 \: V)$. The improved performance is due to the absence of the ground-state absorption peak and an additional component to the Stark shift. Measurements indicate that p-doping the $QD$ barriers can lead to more than a $3x$  increase in $FoM$ modulator performance between temperatures of $-73 \: ^{\circ}C$ to $100 \: ^{\circ}C$ when compared with the stack with $NID$ $QD$ barriers.
\end{abstract}

\section{Introduction}
Nowadays, there have been significant advancements in integrating $InAs/In(Ga)As$ $QD$ lasers into the silicon photonics platform \cite{Norman2018,202000037} with little degradation in device performance. The degradation resilience is due to their confined nature and small size that leads to a lower chance of being affected by defects, including threading dislocations\cite{Norman2018}. $QD$ devices have this advantage over competing technologies, particularly quantum wells ($QWs$). Furthermore, $QD$ lasers offer the prospect of greater temperature tolerance \cite{8840542}, which is fundamentally due to the large energy spacing of conduction states found within the dots. The larger energy spacing prevents any carriers from getting thermally excited to higher states \cite{5943701} and reduces wavelength drift in the laser.

Most research on $QD$ devices has focused on developing lasers over a native substrate and $Si$. With the recent improvements of $QD$ lasers over $Si$ showing relevant results, the focus has shifted to additional devices, including modulators and photodetectors. Despite the prospect of more efficient additional $QD$ devices over $Si$, there have not been many investigations into developing $QD$ modulators. This paper will explore the use of $QDs$ for optical modulation.

The quantum-confined Stark effect ($QCSE$) is widely used in several electro-absorption modulators ($EAMs$) and it is present in $QDs$. There are numerous advantages to utilizing this effect, including faster bandwidth when compared to the plasma dispersion effect \cite{5071309} exploited in current $Si$ modulators. Additionally, the $QCSE$ allows a shorter modulator length and reduced electrical power consumption \cite{vivien2016handbook,Abadia2014}.

The $QCSE$ is present in both $QDs$ and $QWs$ \cite{LIN,Sobhani,Luo_2006,doi:10.1063/1.3119186,1512276,Sandall2013EvaluationOI,Chu:07,Tang:12,6324086}. One of the advantages of $QDs$ devices is that they are more resilient than $QW$ devices when growing them over $Si$ since the $QWs$ are always affected by defects arising from the material lattice mismatch and different thermal expansion coefficients. Additionally, defects in the semiconductor materials lead to uneven static electric fields across $QW$ active regions through the onset of leakage current as demonstrated in \cite{Edwards:13,doi:10.1063/1.118774}. Furthermore, the non-uniform static electric field leads to a non-uniform Stark shift, broadening the $QCSE$ excitonic peak and reducing the maximum absorption, leading to smaller extinction ratios ($ERs$) \cite{Edwards:13,doi:10.1063/1.118774}.

On the other hand, $QDs$ offer the prospect of significant electro-optic coefficients and steeper absorption edges due to their confined nature \cite{Norman2018}. For example, the work in \cite{LIN} demonstrated that the $QCSE$ in $QDs$ can achieve comparable $ERs$ to $QWs$ despite having a lower density of states by two orders of magnitude. This result highlighted the strong $QCSE$ in $QD$ structures due to their increased carrier confinement and suggested their potential for high-performance $EAMs$. 

There have been several measurements of the $QCSE$ in $QDs$ \cite{Chu:07,doi:10.1063/1.3119186,LIN,Sobhani}. Nevertheless, the $QD$ stacks used in those works are very similar to the $QD$ laser, and the stacks were not optimized to improve the $QCSE$ for modulation.

This is not the case of the $QW$ stacks, which were thoroughly optimized to improve the $QCSE$ for modulation. One successful method is the incorporation of barrier doping as shown in \cite{1512276}. The study demonstrated that incorporating a modulation-doped superlattice in a multi-$QW$ structure can lead to equivalent $ERs$ while reducing the applied voltage by almost half. The enhancement was attributed to the modification of the potential by the doping profile to create higher polarizability of the electron and hole wavefunctions. This improvement would lead to more efficient $QW$ modulators with reduced power consumption \cite{Millerenergy}.

The same technique was applied to the $QD$ laser by doping the $QD$ barriers and offering better performance as described in \cite{Zhang2018}. Nevertheless, the method has not been investigated for modulation, and there is no report on the influence of $QD$ barrier doping on the $QCSE$.

In this paper, we investigate the influence of p-doping the $QD$ barriers on the $QCSE$ and compare it with non-intentionally doped ($NID$) barriers with temperatures between  $-73 \: ^{\circ}C$ and $100 \: ^{\circ}C$. The $QCSE$ in both stacks were measured, and the stacks were compared for modulation using the $FoM=\Delta\alpha/\alpha(1 \: V)$, where $\Delta\alpha$ is the change in absorption for a given reverse bias voltage swing and $\alpha(1 \: V)$ is the absorption under an applied reverse bias of $1 \: V$ \cite{Feng2011,Srinivasan2020}. The measurements show that the p-doping in the QD barriers offers up to $3x$ larger $FoM$ between $-73 \: ^{\circ}C$ and $100 \: ^{\circ}C$.

\section{Comparison of the quantum-confined Stark effect with p-doped and non-intentionally doped quantum dot barriers}

This work measured the $QCSE$ in the stacks shown in Table 1 and compared them for modulation. The only difference between the stacks is that the $QD$ barrier is p-doped ($PD$) or left $NID$; the barrier is highlighted in bold in Table 1. This work considers that a $NID$ semiconductor has a doping concentration below $<1 \cdot 10^{16} \: cm^{-3}$. Additionally, we compared the performance from $-73 \: ^{\circ}C$ to $100 \: ^{\circ}C$ by using a heater and a heat sink.

\begin{table}[htbp]
\centering
\arrayrulecolor{black}
\small
\renewcommand{\arraystretch}{.6}
\begin{tabular}{| c  c  c  c | }
	\hline 
    \textbf{Material} & \textbf{Thickness} & \textbf{Doping} & \textbf{Repetitions} \\
    \hline
    $GaAs$ & $300 \: nm$ & $P$ ($10^{19} \: cm^{-3}$) & $1 \times$\\
    \hline
    $GaAs/Al_{0.4}Ga_{0.6}As$ & $5 \: nm$ & $P$ ($5 \cdot 10^{17} \: cm^{-3}$) & $10 \times$\\
    \hline
    $Al_{0.4}Ga_{0.6}As$ & $1400 \: nm$ & $P$ ($2 \cdot 10^{17} \: cm^{-3}$) & $1 \times$\\
    $Al_{0.4}Ga_{0.6}As$ & $30 \: nm$ & $NID$ & $1 \times$\\
    \hline
    $GaAs/Al_{0.4}Ga_{0.6}As$ & $2 \: nm$ & $NID$ & $12 \times$\\
    \hline
    $GaAs$ & $17.5 \: nm$ & $NID$ & \\
    \textbf{GaAs} & \textbf{10 \:  nm} & \textbf{P ($5 \cdot 10^{17} \: cm^{-3}$)} & \\
      &   & \textbf{or NID ($<1 \cdot 10^{16} \: cm^{-3}$)} &
    \\
    $GaAs$ & $10 \: nm$ & $NID$ & \\
    $GaAs$ & $5 \: nm$ & $NID$ & $7 \times$ \\
    $In_{0.16}Ga_{0.84}As$ & $5 \: nm$ & $NID$ & \\
    $InAs$ $QD$ & $3 \:  ML$ & $NID$ & \\
    $In_{0.16}Ga_{0.84}As$ & $2 \:  nm$ & $NID$ & \\
    \hline
    $GaAs$ & $42.5 \:  nm$ & $NID$ & $1 \times$\\
    \hline
    $GaAs/Al_{0.4}Ga_{0.6}As$ & $2 \: nm$ & $NID$ & $12 \times$ \\
    \hline
    $Al_{0.4}Ga_{0.6}As$ & $30 \: nm$ & $NID$ & $1 \times$ \\
    $Al_{0.4}Ga_{0.6}As$ & $1400 \: nm$ & $N$ ($2 \cdot 10^{18} \: cm^{-3}$) & $1 \times$\\
    \hline
    $GaAs/Al_{0.4}Ga_{0.6}As$ & $5 \: nm$ & $N$ ($2 \cdot 10^{18} \: cm^{-3}$) & $10 \times$\\
    \hline
    $GaAs$ & $200 \: nm$ & $N$ ($2 \cdot 10^{18} \: cm^{-3}$) & $1 \times$\\
    \hline
\end{tabular}
\caption{Layer structure for $QD$ stacks. The only difference between the stacks is the $QD$ barrier doping highlighted in bold, which is p-doped ($5 \cdot 10^{17} \: cm^{-3}$) or left $NID$ ($<1 \cdot 10^{16} \: cm^{-3}$). $P$ stands for p-doped, and $N$ for n-doped. The column $Repetitions$ represents the number of times a set of layers is repeated in the stack, e.g., there are 7x repetitions of the $QD$ active region.}
\label{Table:table1}
\end{table}

The stacks in Table 1 were grown using molecular beam epitaxy on an n-doped $GaAs$ substrate. The only difference between the stacks is the $10 \: nm$ GaAs $QD$ barrier layer highlighted in bold, which is either p-doped ($5 \cdot 10^{17} \: cm^{-3}$) or left $NID$ ($<1 \cdot 10^{16} \: cm^{-3}$). The p-doping concentration in the $QD$ barrier of $5 \cdot 10^{17} \: cm^{-3}$ is equivalent to $10$ holes per dot. In both stacks, the $InAs$ $QD$ areal density is $5 \cdot 10^{10} \: cm^{-2}$. The $QD$ active region stack consists of 7 dot-in-a-well layers with a total thickness of $350 \: nm$. Finally, the waveguide core has an $Al_{0.4}Ga_{0.6}As$ cladding layer above and below with a thickness of $1400 \: nm$. The structure contains highly p-doped top and n-doped bottom $GaAs$ layers for making electrical contacts. To handle strain between $GaAs$ and $Al_{0.4}Ga_{0.6}As$, there are $10$ repetitions of $5 \: nm$ $GaAs/Al_{0.4}Ga_{0.6}As$.

\begin{figure}[H]
\centering
\subfloat{{\includegraphics[width=8.5cm]{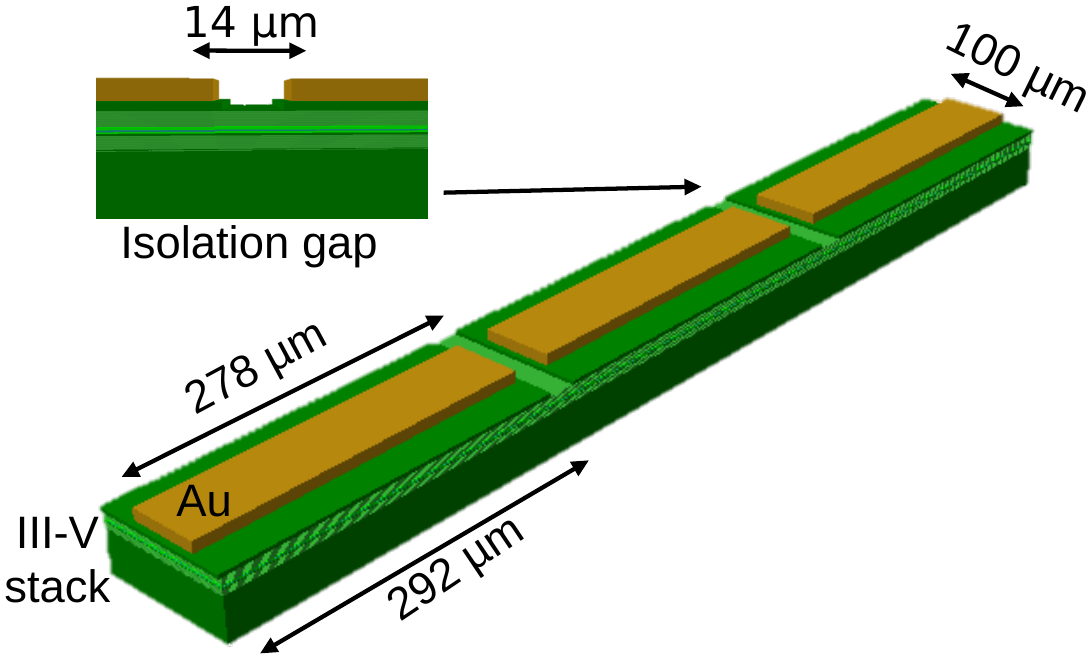}}}
\caption{\label{fig:figure1}Schematic of multisection device used to measure the $QCSE$ in the stacks of Table 1.}
\end{figure}

The $QCSE$ was measured using the segmented contact method \cite{Mahoney:21} using the structure shown in Fig. \ref{fig:figure1}. Two identical structures were fabricated with the stacks shown in Table 1. The top gold ($Au$) contacts have dimensions $100 \: \mu m \times 278 \: \mu m$, and the separation between them is $14 \: \mu m$. The top layer in the stack is doped over $1 \cdot 10^{19} \: cm^{-3}$ to guarantee Ohmic contacts with $Au$. Gaps of $14 \: \mu m$ in length and $700 \: nm$ in depth were etched in the top layers to isolate the contacts electrically. During the measurement, the leakage current was compensated as the resistance between contacts and the leakage current change with temperature from $-73 \: ^{\circ}C$ to $100 \: ^{\circ}C$.


\begin{figure}[htbp]
\centering
\begin{minipage}{.5\textwidth}
  \centering\includegraphics[width=1\linewidth]{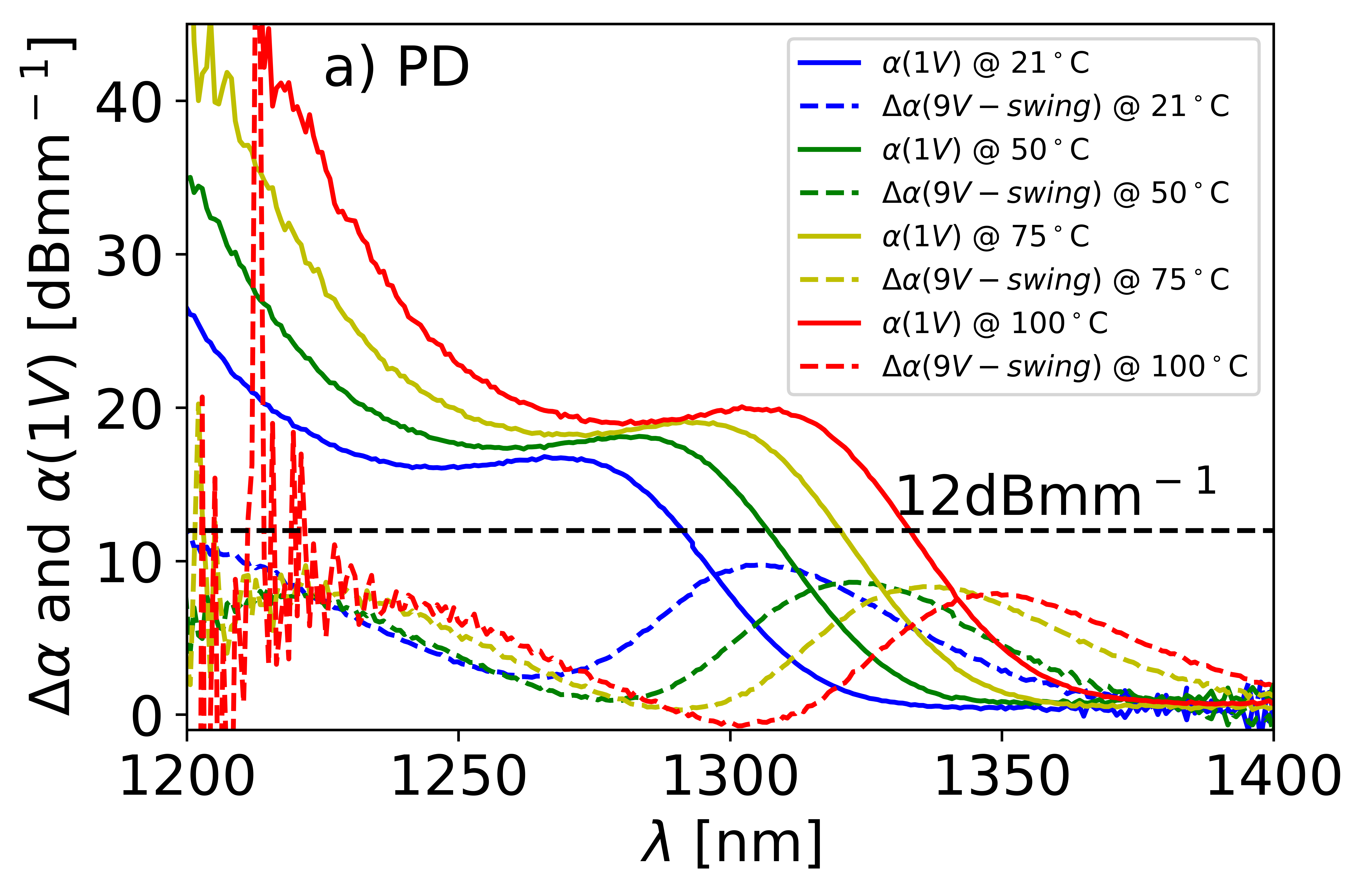}
\end{minipage}%
\begin{minipage}{.5\textwidth}
  \centering\includegraphics[width=1\linewidth]{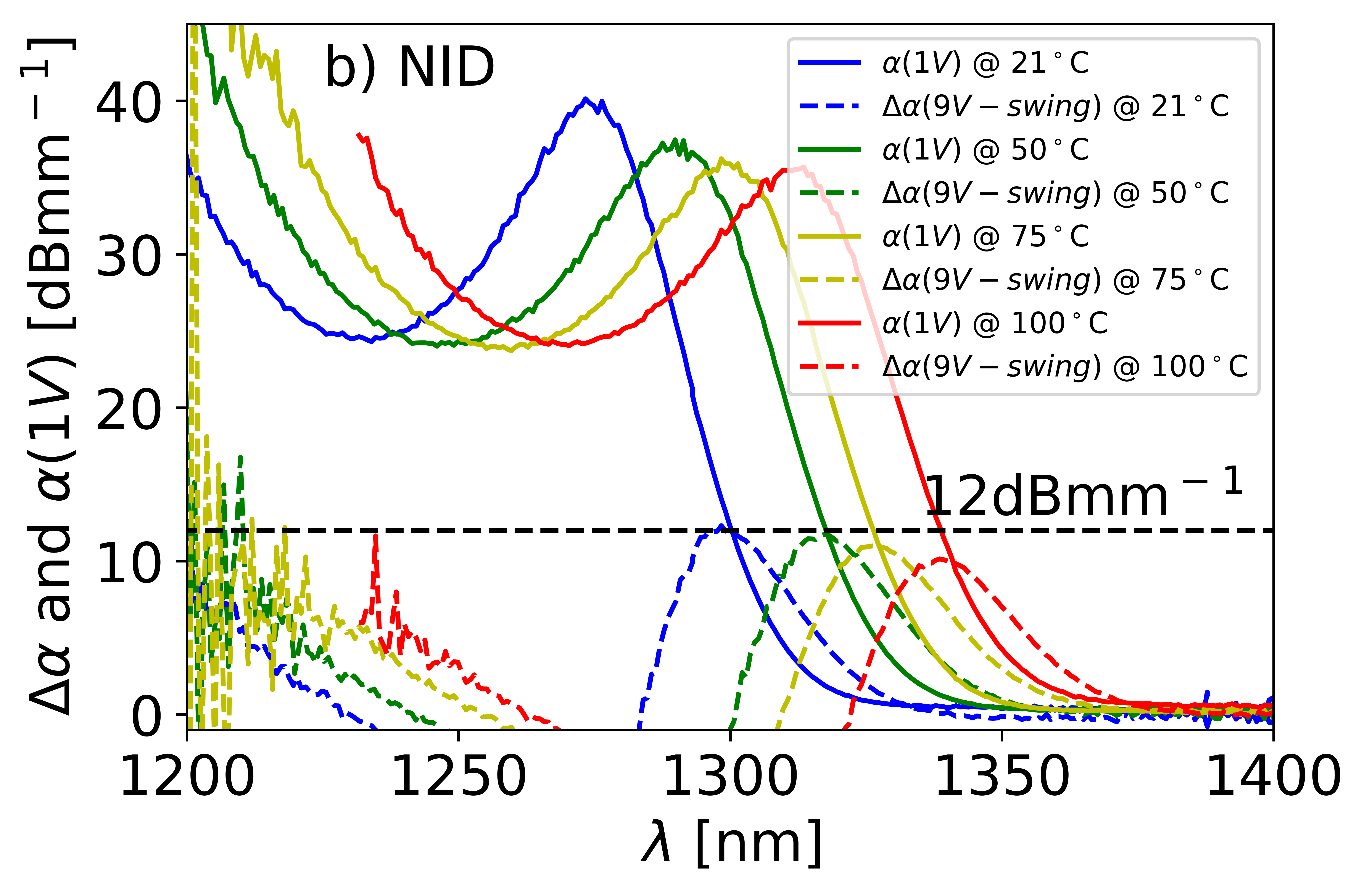}
\end{minipage}
\caption{\label{fig:figure2} Change in absorption $\Delta\alpha$ for a voltage swing of $9 \: V$ and absorption at a pre-biased value of $1 \: V$ $\alpha(1 \: V)$ for p-doped ($PD$) (a) and $NID$ (b) $QD$ barriers for $21 \: ^{\circ}C$, $50 \: ^{\circ}C$, $75 \: ^{\circ}C$, and $100 \: ^{\circ}C$.}
\end{figure}

The measurements of the $QCSE$ are shown in Fig. \ref{fig:figure2}(a) for the stack with doped $QD$ barriers and Fig. \ref{fig:figure2}(b) for the stack with $NID$ $QD$ barriers. Both figures show the change in absorption $\Delta\alpha$ for a reverse bias voltage swing of $9 \: V$ and the absorption $\alpha(1 \: V)$ at a reverse bias of $1 \: V$ for $21 \: ^{\circ}C$, $50 \: ^{\circ}C$, $75 \: ^{\circ}C$, and $100 \: ^{\circ}C$. The pre-biased value of $1 \: V$ is used to mitigate the modulator's capacitance penalty as in \cite{Sobhani}.



When comparing the $\Delta\alpha$ of both stacks in Fig. \ref{fig:figure2}(a) and (b), there is a slight decrease in $\Delta\alpha$ with temperature. This will lead to a modulator with a stable $ER$ from $21 \: ^{\circ}C$ to $100 \: ^{\circ}C$ for both stacks. Additionally, the stack with $NID$ $QD$ barriers has the largest $\Delta\alpha$ value of around $12 \: dBmm^{-1}$, which will lead to a larger $ER$. Nevertheless, it also has a larger material absorption $\alpha(1 \:V)$, which will lead to a larger insertion loss ($IL$).

Regarding $\alpha(1 \: V)$ shown in Fig. \ref{fig:figure2}(a), the stack with barrier doping does not exhibit an absorption peak at the $QD$ ground-state wavelength. The ground-state absorption peak suppression is attributed to the injection of holes in the $QD$ valence ground-state from the p-doped barrier. The holes inhibit the excitation of electrons from the valence ground-state to the conduction ground-state and inhibit photon absorption. Furthermore, the doping in the barriers will contribute to increasing the absorption. Nevertheless, the contribution is minimal due to the small area of the barriers compared to the entire active region. On the other hand, the stack with $NID$ barriers has the expected ground-state absorption peak, as shown in Fig. \ref{fig:figure2}(b).

Additionally, the maximum absorption $\alpha(1 \: V)$ around $1275 \: nm$ of the stack with doped barriers increases slightly with increasing temperature, as shown in Fig. \ref{fig:figure2}(a). This behavior is attributed to the redistribution of carriers in the valence ground-states as the temperature increases. The redistribution leads to a greater occupancy probability for electrons in the top $QD$ valence states, which increases photon absorption. On the other hand, the ground-state absorption peak of the stack with $NID$ barriers in Fig. \ref{fig:figure2}(b) decreases with increasing temperature as expected from the $QCSE$ in $QDs$. The same trend was measured in \cite{Cornet2004}.

\begin{figure}[htbp]
\centering
\begin{minipage}{.5\textwidth}
  \centering\includegraphics[width=1\linewidth]{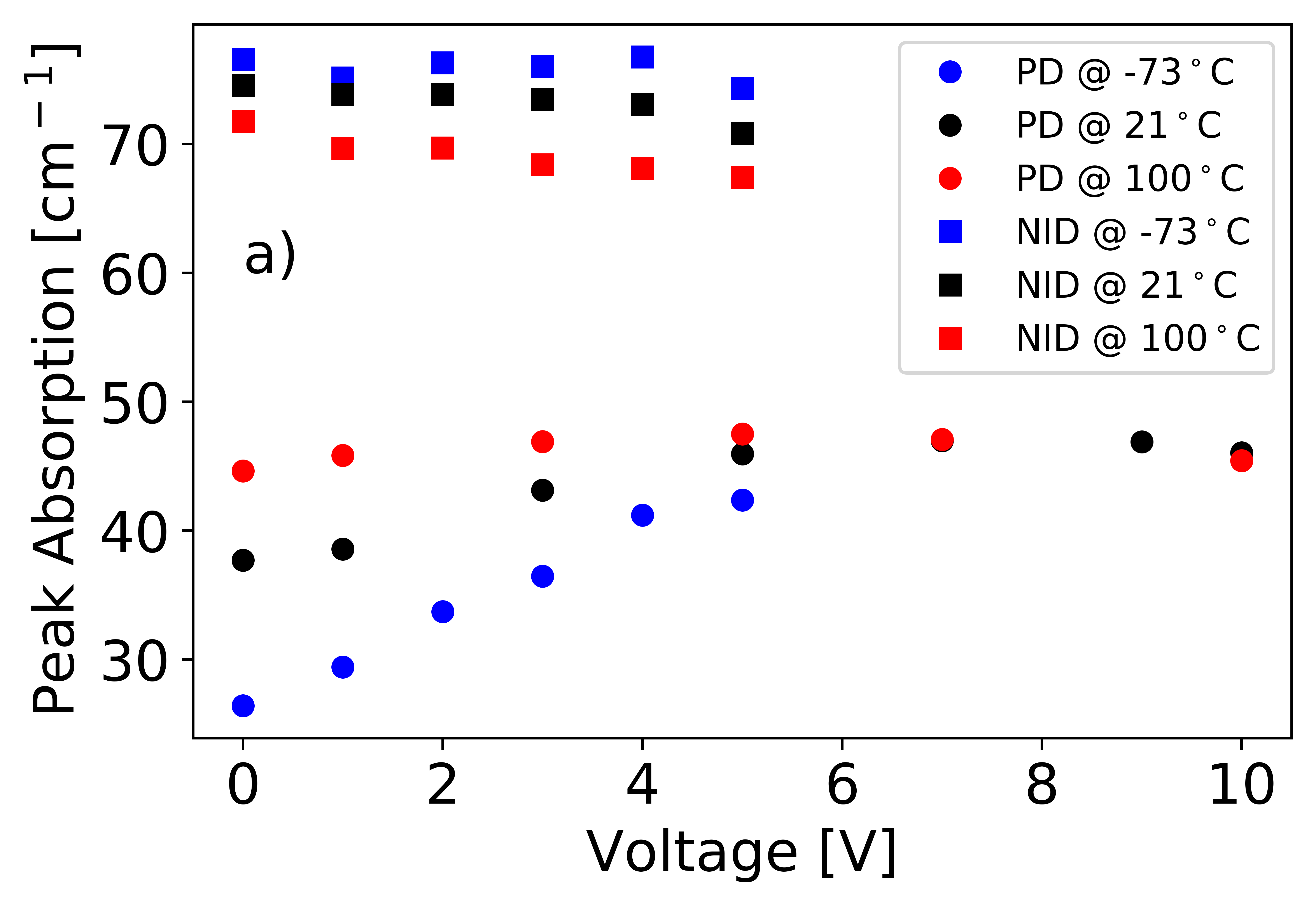}
\end{minipage}%
\begin{minipage}{.5\textwidth}
  \centering\includegraphics[width=1.05\linewidth]{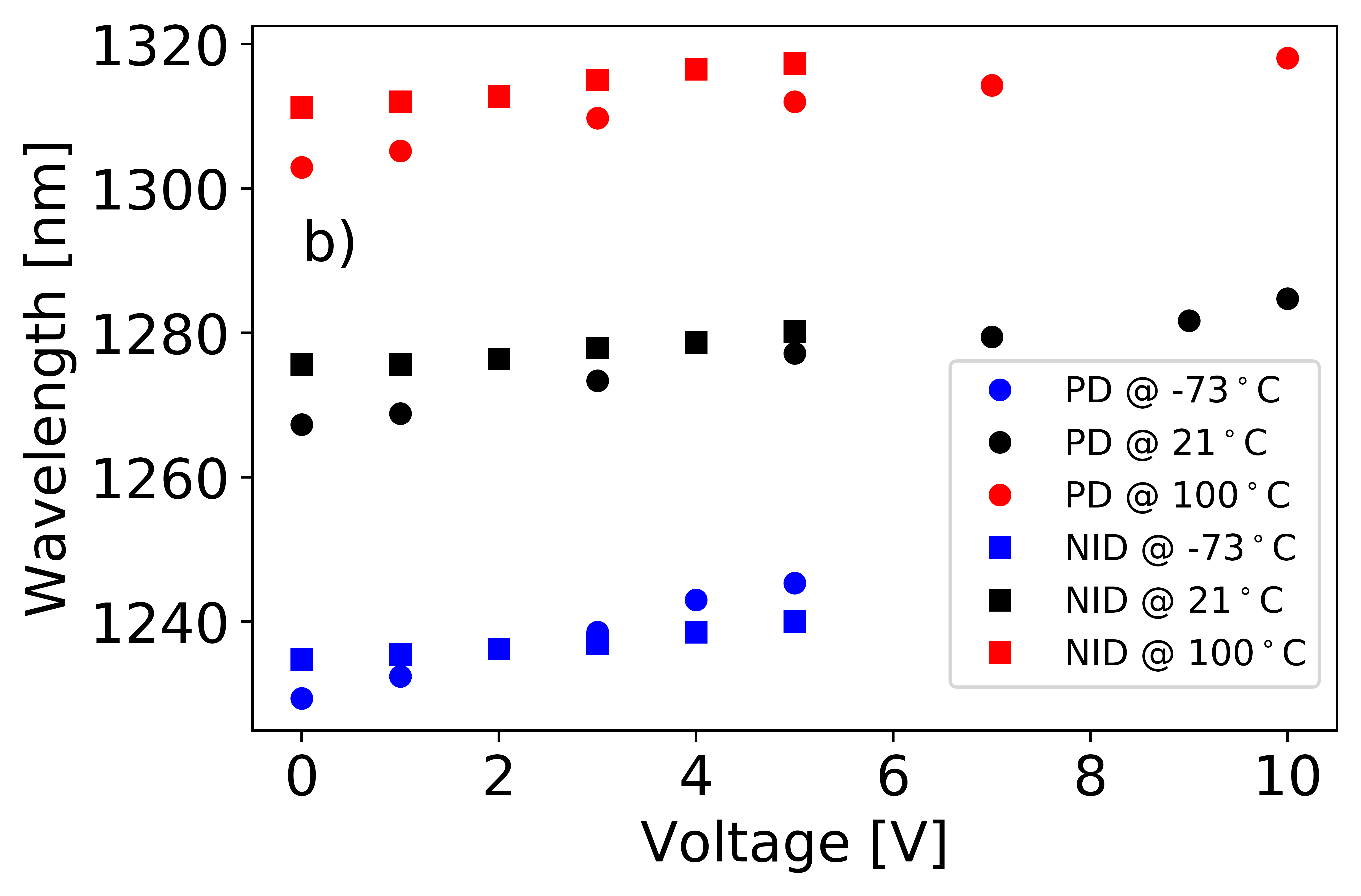}
\end{minipage}
\caption{\label{fig:figure3} The change in ground-state absorption peak (a) and absorption peak wavelength (b) for $-73 \: ^{\circ}C$, $21 \: ^{\circ}C$, and  $100 \: ^{\circ}C$ for the stack with p-doped ($PD$) and $NID$ barriers.}
\end{figure}

To investigate the trends further, the $QCSE$ ground-state absorption peak for the stack with p-doped ($PD$) and $NID$ $QD$ barriers as a function of reverse bias was measured in Fig. \ref{fig:figure3}(a) for three temperatures $-73 \: ^{\circ}C$, $21 \: ^{\circ}C$, and $100 \: ^{\circ}C$. Additionally, the absorption peak wavelength shift is represented in Fig. \ref{fig:figure3}(b). The ground-state absorption peak wavelength is the wavelength that has the ground-state absorption peak.


As shown in Fig. \ref{fig:figure3}(a), the stack with $NID$ barriers has a constant reduction in the ground-state absorption peak with increasing reverse bias at $-73 \: ^{\circ}C$, $21 \: ^{\circ}C$, and $100 \: ^{\circ}C$. As the reverse bias increases, the increased static electric field pulls the electrons and holes in opposite directions inside the $QDs$. This phenomenon leads to a reduction in the overlap between the carriers' wavefunctions and contributes to reducing the photon absorption \cite{PhysRevLett.84.733}.
 
On the other hand, the stack with p-doped $QD$ barriers has different behavior. Below moderate reverse bias (< $5.5 \: V$), the holes at the valence state's top inhibit the photon absorption. As the reverse bias increases, the built-in static electric field depletes the holes from the top valence state, allowing photons to excite electrons from the valence ground-state to the bottom of the conduction ground-state. For voltages below $4 \: V$, the ground-state absorption peak is smaller at $-73 \: ^{\circ}C$ rather than at $100 \: ^{\circ}C$ due to more electrons being thermally excited to the top of the valence ground-state at higher temperatures. For the same reason, the gradient of the ground-state absorption peak at $-73 \: ^{\circ}C$ is more significant between $0 \: V$ and $5.5 \: V$.
 
The ground-state absorption peak will increase with a reverse bias for all temperatures until all holes are entirely depleted from the top of the valence ground-state. As shown in Fig. \ref{fig:figure3}(a), all holes are depleted around $5.5 \: V$, and the absorption peak starts to decrease beyond $5.5 \: V$ for the same reason as the stack with $NID$ barriers. The built-in static electric field pulls the carriers in opposite directions reducing the carrier's wavefunction overlap and the photon absorption.

The wavelength shift of the ground-state absorption peak is shown in Fig. \ref{fig:figure3}(b). The stack with doped barriers shows a more significant shift than the stack with $NID$ barriers. For reverse biases between $1 \: V$ and $5 \: V$, red-shifts at $-73 \: ^{\circ}C$ are $13 \: nm$ for the stack with doped barriers and $5 \: nm$ for the stack with $NID$ barriers. The shift difference is more considerable at $-73 \: ^{\circ}C$, and it decreases with increasing temperature.

\begin{figure}[H]
\centering\includegraphics[width=0.575\textwidth]{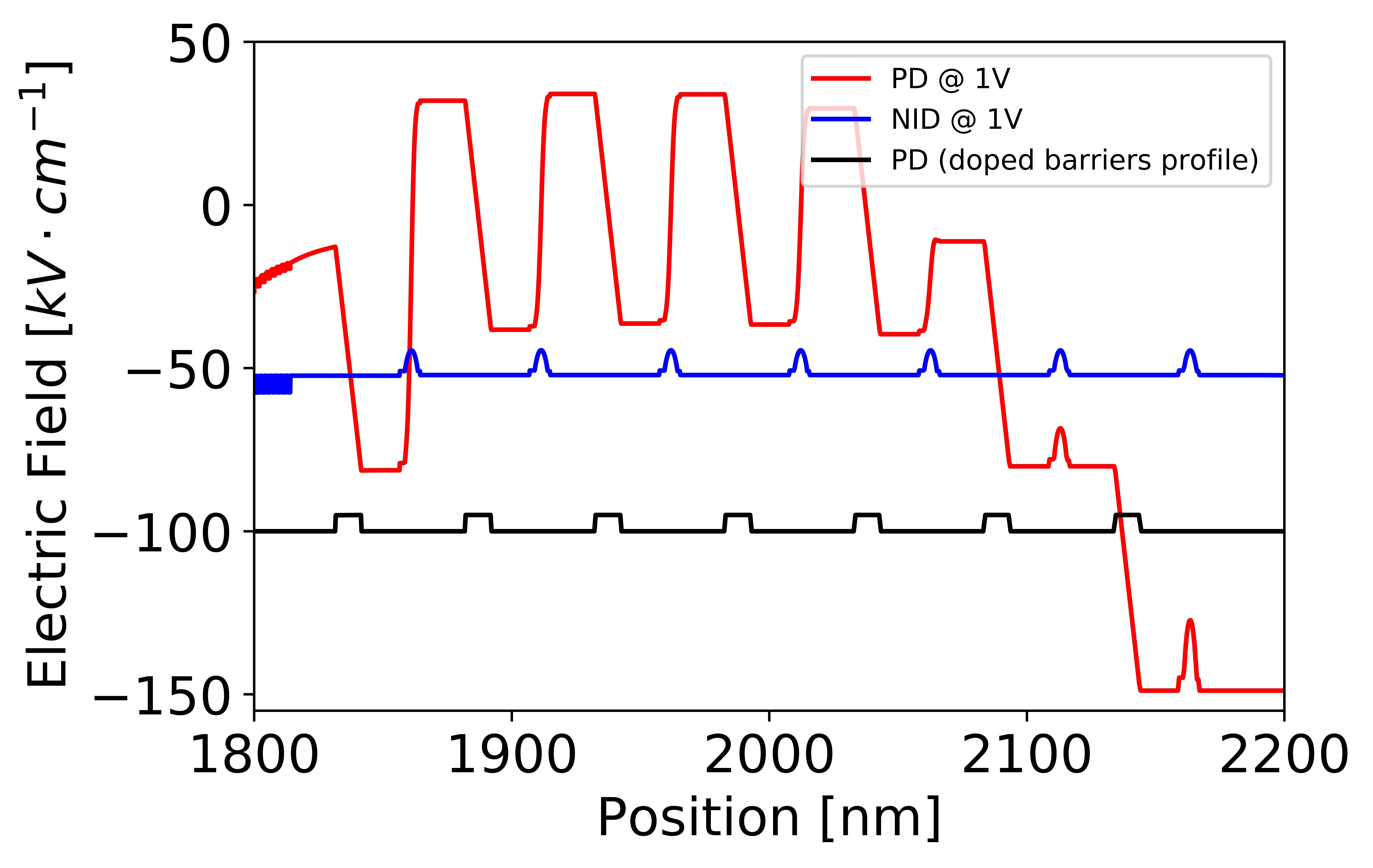}
\caption{\label{fig:figure4} Static electric field distribution across the $QD$ active region for the stacks with doped (red line) and $NID$ (blue line) barriers for a $1 \: V$ reverse bias at $21 \: ^{\circ}C$. The black line shows the position of the doped barriers, and the $QDs$ are between them. The static electric field was numerically calculated using nextnano \cite{NN}.}
\end{figure}

Both stacks show the Stark shift present in the $QCSE$. Nevertheless, the stack with doped barriers has additional contributions to the red-shift. One comes from the $QD$ layers having a larger static electric field strength when compared with the stack with $NID$ barriers. The larger static electric field  is due to the $QD$ layers being surrounded by doped $GaAs$ layers. The static electric field distribution across the $QD$ active region for both stacks is shown in Fig. \ref{fig:figure4} and it was calculated using nextnano \cite{NN}. 

The black line at the bottom of the figure indicates the position of the doped barriers. Figure \ref{fig:figure4} highlights the significantly stronger field across the $QDs$ of the stack with doped barriers. Therefore, the larger static electric field would cause a more substantial red-shift.

The other shift contribution originates from the removal of holes from the highest valence states as the static electric field increases with reverse bias. As holes are depleted from the top of the valence states, longer wavelength (lower energy) transitions are allowed.

\begin{figure}[htbp]
\centering
\begin{minipage}{.5\textwidth}
  \centering
  \includegraphics[width=1\linewidth]{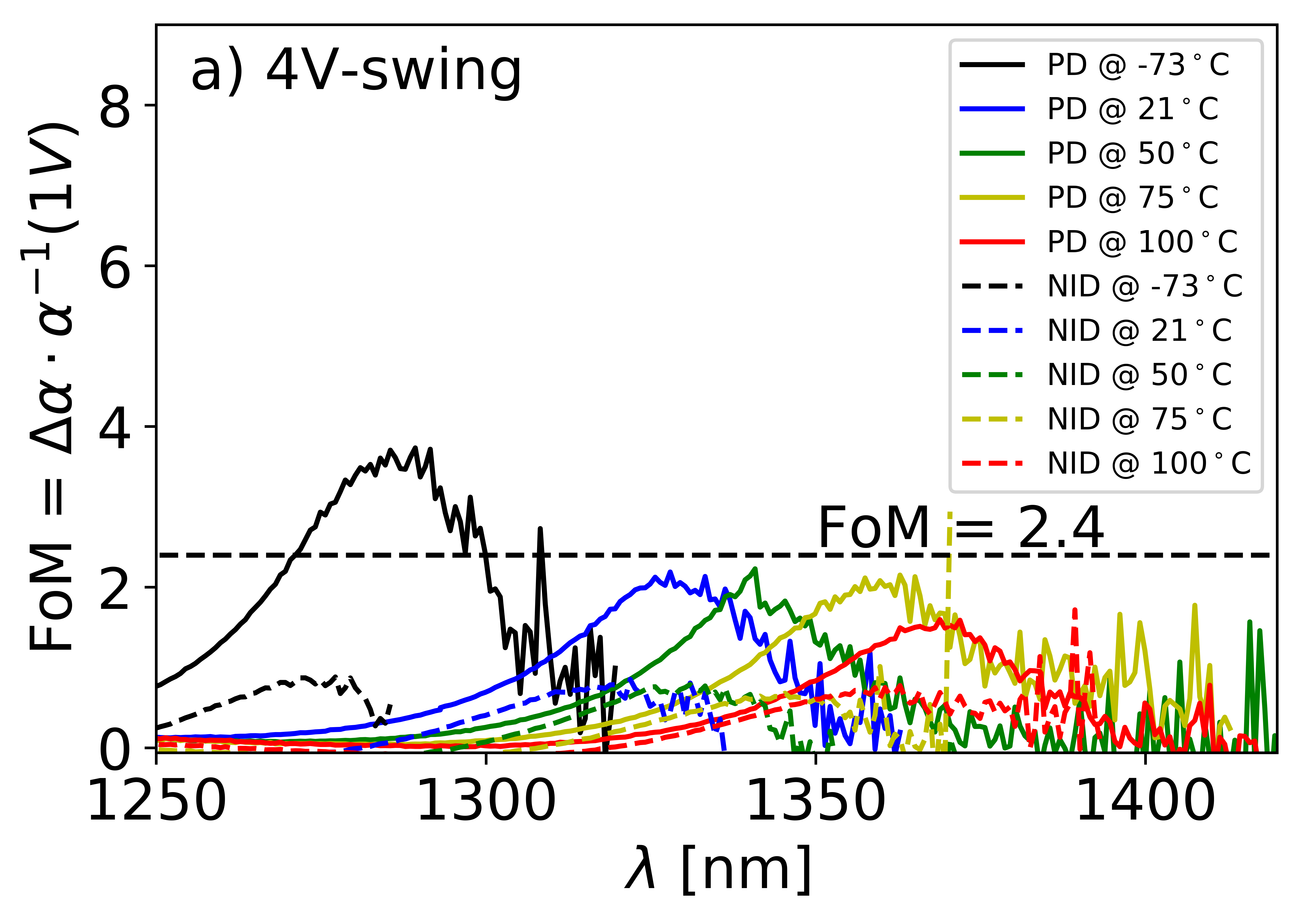}
\end{minipage}%
\begin{minipage}{.5\textwidth}
  \centering
  \includegraphics[width=1\linewidth]{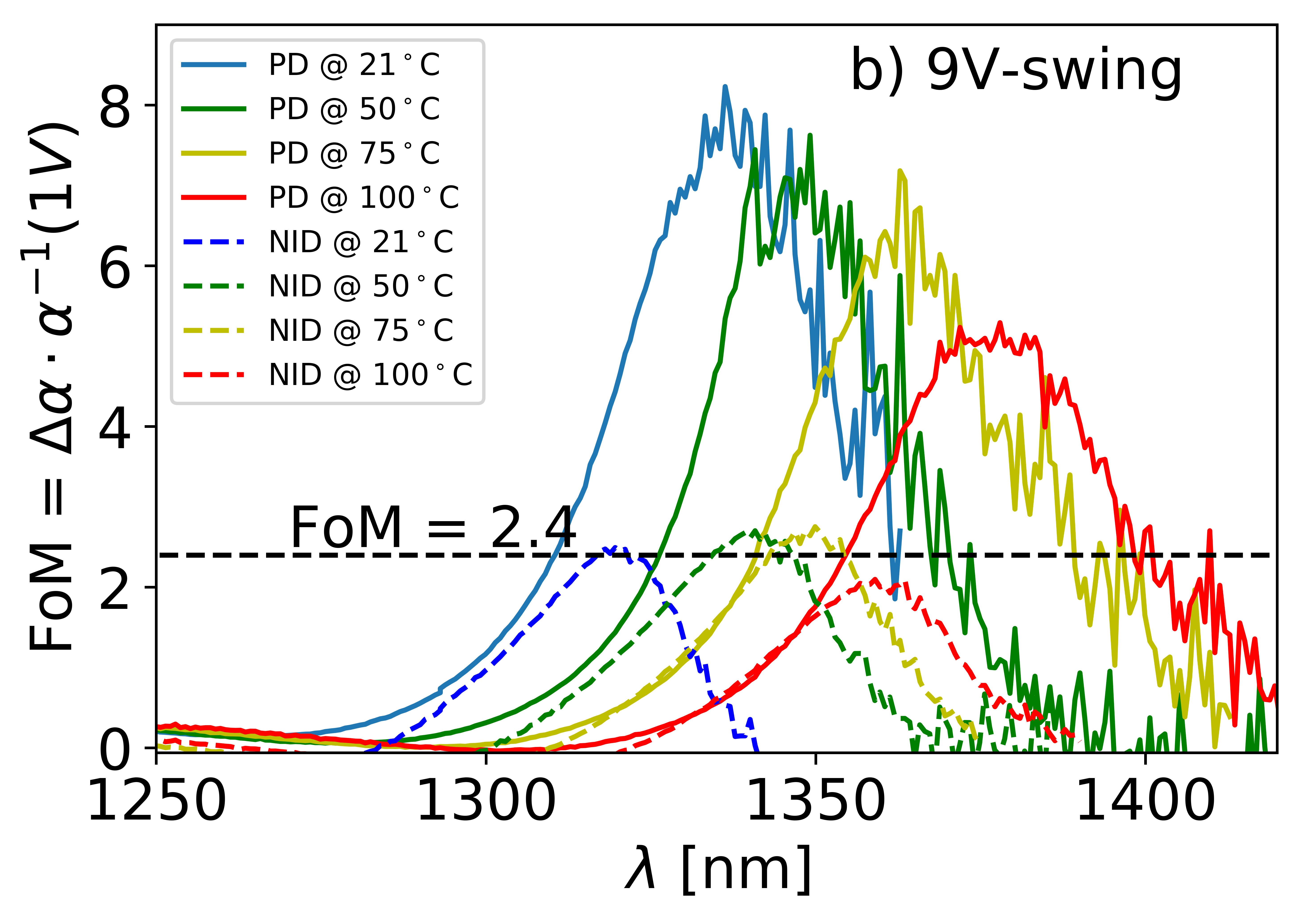}
\end{minipage}
\caption{\label{fig:figure5}$FoM$ for a $4 \: V$-swing (a) and $9 \: V$-swing (b) for both stacks with p-doped and $NID$ barrier vs temperature.}
\end{figure}

In order to determine which stack design is better for modulation, the stacks were compared using the $FoM=\Delta\alpha$/$\alpha(1 \: V)$. The $FoM$ maximizes $\Delta\alpha$ to have a large $ER$ and minimizes $\alpha(1 \: V)$ to reduce the $IL$. The $FoM$ is shown in Fig. \ref{fig:figure5}(a) for a $4 \: V$-swing and in Fig. \ref{fig:figure5}(b) for a $9 \: V$-swing. A maximum swing of $9 \: V$ was used to ensure working below the breakdown static electric field of the materials. As shown in Fig. \ref{fig:figure5}, the stack with p-doped $QD$ barriers outperforms the stack with $NID$ barriers offering up to $3x$ enhancement in the $FoM$ from $-73 \: ^{\circ}C$ to $100 \: ^{\circ}C$.


Several factors explain the better performance of the stack with doped $QD$ barriers. As shown in Fig. \ref{fig:figure2}, the $\Delta\alpha$ of the stack with $NID$ barriers is around $12 \: dBmm^{-1}$, and it is slightly larger than the stack with doped barriers. Nevertheless, the $\alpha(1 \: V$) is more significant in the stack with $NID$ barriers. In this comparison, we only considered the material absorption contribution to the $IL$ since both stacks will have a similar loss due to wall roughness in the modulator's waveguide.

Additionally, the extra red-shift in the stack with doped barriers shifts away the maximum $\Delta\alpha$ from the absorbing edge of the absorption spectrum $\alpha(1\: V$) represented in Fig. \ref{fig:figure2}(a) offering a better trade-off $\Delta\alpha$/$\alpha(1 \: V)$. Overall, the $FoM$ is better for the stack with doped barriers when considering all factors.

To explain the results in Fig. \ref{fig:figure5}(a) further, the $FoM$ of the stack with $NID$ barriers is below $\sim0.6-0.9$ from $-73 \: ^{\circ}C$ to $100 \: ^{\circ}C$ for a $4 \: V$-swing. On the other hand, the $FoM$ of the stack with doped barriers is $\sim4$ at $-73 \: ^{\circ}C$ and it is around $\sim2.4$ from $21 \: ^{\circ}C$ to $100 \: ^{\circ}C$. Consequently, for a $4 \: V$-swing the $FoM$ of the stack with doped barriers is $\sim4.5x$ times larger than the stack with $NID$ barriers.

Considering the results for a $9 \: V$-swing shown in Fig. \ref{fig:figure5}(b), the maximum $FoM$ of the stack with $NID$ barriers is around $2.4$ from $21 \: ^{\circ}C$ to $100 \: ^{\circ}C$. On the other hand, the maximum $FoM$ for the stack with doped barriers starts at $\sim8$ at $21 \: ^{\circ}C$ and degrades to $\sim6$ at $100 \: ^{\circ}C$. Consequently, the $FoM$ of the stack with doped barriers is at least $\sim3x$ (at $100 \: ^{\circ}C$) times larger than the stack with $NID$ barriers. Finally, the $FoM=2.4$ of the stack with $NID$ barriers at $9 \: V$-swing is the same as the stack with doped barriers with a $4 \: V$-swing. This result implies that a modulator using the stack with doped barriers may need $\sim 5 \: V$ less driving voltage which would significantly reduce power consumption.

\section{Comparison with the state-of-the-art}

The $QCSE$ was measured in $QDs$ in slightly different stacks. Using different stack configurations can lead to changes in the $QCSE$, which will produce modulators with various performances. All the papers measuring the $QCSE$ in $QDs$ are summarized in Table 2.

Several data/telecom applications require a minimum $ER$ of around $20 \: dB$. For the mentioned $ER$, the sample with $NID$ barriers will have a length of about $2.6 \: mm$ and a material absorption contribution to the $IL$ around $10.3 \: dB$. For most applications, that $IL$ results in most optical interconnects going over the power budget. On the other hand, the sample with doped barriers will require $2.2 \: mm$. The material absorption contribution to the $IL$ will be around $5.6 \: dB$, which will not exceed most optical interconnects power budgets.

This comparison only considers the material loss contribution to the $IL$. To compare both design stacks, all $ER$ and $IL$ were calculated assuming a modulator length of $1 \: mm$ to obtain an $ER \sim 10 \: dB$.

\begin{table}[htbp]
 \centering \caption{\label{tab:table1} Measurements of the $QCSE$ in $QDs$ with different stacks. The values of the $ERs$ and the $ILs$ were calculated using a length of $1 \: mm$ to obtain $ERs$ around $10 \: dB$. The values marked with a -- (dash) are not available. $V_{reverse}$ stands for the reverse bias voltage, $BW$ for the electrical bandwidth, and $\lambda$ for the wavelength.}
\arrayrulecolor{black}{}
\begin{tabular}{p{1cm}|p{1.4cm}p{1.4cm}p{1.4cm}p{1.4cm}p{1.4cm}p{1.4cm}}
\hhline{>{\arrayrulecolor{black}}}
     & Cambridge \cite{Chu:07} (2007) & NTU \cite{doi:10.1063/1.3119186} (2009) & NSYSU \cite{LIN} (2011) & Glasgow \cite{Sobhani} (2019) &  Our Work [$NID$]${^2}$ & Our Work [$PD$]${^2}$\\
    \hline
    $ER$ & $3.5 \: dB$ & $10 \: dB$ & $\sim10 \: dB{^1}$ & $\sim 12 \: dB$  & $7.6 \: dB$ & $9 \: dB$\\
    $IL$ & -- & -- & --& $\sim3 \: dB$ & $3.9 \: dB$ & $2.5 \: dB$\\
    $V_{reverse}$ & $8 \: V$ & $10 \: V$ & $5 \: V$ & $11 \: V$  & $9 \: V$ & $9 \: V$\\
    $BW$ & $2 \: GHz$ & --& $3.3 \: GHz$ & --  & -- & --\\
    $\lambda$ & $1300 \: nm$ & $1328 \: nm$ & $1300 \: nm$ & $1310 \: nm$  & $1311 \: nm$ & $1316 \: nm$\\
   \end{tabular}
\smallskip
\parbox[t]{\textwidth}{\footnotesize
  $^1$ $ER$ taken from $1 \: V$ in Fig. 4 \cite{LIN} to be consistent with the rest of the papers. \\
  $^2$ The values are estimated at $21 \: ^{\circ}C$.
}
    \end{table}

The work presented in \cite{Chu:07} reported the second-largest bandwidth around $2 \: GHz$ but the $3.5 \: dB \: ER$ for a $8 \: V$-swing is significantly lower than the one presented in this work. On the other hand, the $ER$ achieved in \cite{doi:10.1063/1.3119186} and \cite{LIN} are around $10 \: dB$, which is a similar value to the one presented in this paper. However, stacks have 10 (\cite{doi:10.1063/1.3119186}) and 12 (\cite{LIN}) $QD$ layers. The additional $QD$ layers absorb additional light due to a more overlap between the absorbing $QDs$ and the optical mode. Having a thicker active region will lead to a smaller electrical bandwidth.

Finally, the work presented in \cite{Sobhani} offers similar performance to this work. Nevertheless, it requires a larger driving voltage around $11 \: V$-swing, and the stack has a higher $QD$ density around $5.9 \cdot 10^{10} \: cm^{-2}$ and 8 $QD$ layers.

To conclude, the stack with $NID$ barriers offers similar performance to the state-of-the-art shown in Table 2. Nevertheless, when the barriers are doped, the efficiency of the $QCSE$ for modulation is boosted, offering better performance than the state-of-the-art in terms of $FoM$, driving voltage, electrical power consumption, and operational temperature from $-73\: ^{\circ}C$ to $100 \: ^{\circ}C$.

\section{Conclusion}
This work measured the $QCSE$ in $InAs/In(Ga)As$ $QDs$ using $NID$ and p-doped $QD$ barriers. The measurements indicate that doping $QD$ barriers lead to a $FoM=\Delta\alpha/\alpha(1 \: V)$ at least $3x$ larger than the stack with undoped barriers from $-73 \: ^{\circ}C$ to $100 \: ^{\circ}C$. The better performance is due to the absence of the ground-state absorption peak and an additional component to the Stark shift due to hole depletion in the valence ground-state and a larger static electric field. To conclude, doping barriers offer better performance than the state-of-the-art stacks in terms of $FoM$, driving voltage, electrical power consumption, and operational temperature from $-73\: ^{\circ}C$ to $100 \: ^{\circ}C$.

\section*{Funding}
Engineering and Physical Sciences Research Council (2268875).

\section*{Acknowledgments}
This work is part of the feasibility study project 'Measurement of Carrier-Induced Electro-Refraction and Electro-Absorption in $InAs/In(Ga)As$ Quantum Dots' funded by the EPSRC Future Manufacturing Hubs, including the Compound Semiconductor Manufacturing Hub.

We acknowledge Prof. John Donegan ($Trinity \: College \: Dublin$) for fruitful discussions. We thank Dr. Alireza Samani ($Ciena$), Dr. Md. Ghulam Saber ($Ciena$), Dr. Yannick D'Mello ($McGill \: University$), and Prof. {Nicol\'as} Bouchard ($Intel$) for proofreading this paper.

\section*{Disclosures}
The authors declare no conflicts of interest.

\section*{Data availability statement}
Data underlying the results presented in this paper are not publicly available at this time but may be obtained from the authors upon reasonable request at abadian@cardiff.ac.uk or https://orcid.org/0000-0002-7355-4245.

\bibliography{paperFinalVersionReviewed}

\providecommand{\noopsort}[1]{} \providecommand{\singleletter}[1]{#1}%
\begin{thebibliography}{10}
\newcommand{\enquote}[1]{``#1''}

\bibitem{Norman2018}
J.~C. Norman, D.~Jung, Y.~Wan, and J.~E. Bowers, \enquote{Perspective: The
  future of quantum dot photonic integrated circuits,}
  {\protect\JournalTitle{APL Photonics}} \textbf{3}, 030901 (2018).

\bibitem{202000037}
Y.~Wan, J.~C. Norman, Y.~Tong, M.~J. Kennedy, W.~He, J.~Selvidge, C.~Shang,
  M.~Dumont, A.~Malik, H.~K. Tsang, A.~C. Gossard, and J.~E. Bowers,
  \enquote{1.3 µm quantum dot-distributed feedback lasers directly grown on
  (001) {Si},} {\protect\JournalTitle{Laser \& Photonics Reviews}} \textbf{14},
  2000037 (2020).

\bibitem{8840542}
E.~D. {Le Boulbar}, L.~Jarvis, D.~Hayes, S.~Shutts, Z.~Li, M.~Tang, H.~Liu,
  A.~Samani, P.~M. Smowton, and N.~Abadía, \enquote{Temperature dependent
  behavior of the optical gain and electroabsorption modulation properties of
  an {InAs/GaAs} quantum dot epistructure,} in \emph{2019 21st International
  Conference on Transparent Optical Networks (ICTON),}  (2019), p. 1–4.

\bibitem{5943701}
T.~Kageyama, K.~Nishi, M.~Yamaguchi, R.~Mochida, Y.~Maeda, K.~Takemasa,
  Y.~Tanaka, T.~Yamamoto, M.~Sugawara, and Y.~Arakawa, \enquote{Extremely high
  temperature (220°c) continuous-wave operation of 1300-nm-range quantum-dot
  lasers,} in \emph{2011 Conference on Lasers and Electro-Optics Europe and
  12th European Quantum Electronics Conference (CLEO EUROPE/EQEC),}  (2011), p.
  1–1.

\bibitem{5071309}
D.~A.~B. Miller, \enquote{Device requirements for optical interconnects to
  silicon chips,} {\protect\JournalTitle{Proceedings of the IEEE}} \textbf{97},
  1166–1185 (2009).

\bibitem{vivien2016handbook}
L.~Vivien and L.~Pavesi, \emph{Handbook of Silicon Photonics}, Series in Optics
  and Optoelectronics (CRC Press, 2016).

\bibitem{Abadia2014}
N.~Abadía, T.~Bernadin, P.~Chaisakul, S.~Olivier, D.~Marris-Morini, R.~E.
  de~Lamaëstre, J.~C. Weeber, and L.~Vivien, \enquote{Low-power consumption
  {Franz-Keldysh} effect plasmonic modulator,} {\protect\JournalTitle{Opt.
  Express}} \textbf{22}, 11236–11243 (2014).

\bibitem{LIN}
C.-H. Lin, J.-P. Wu, Y.-Z. Kuo, Y.-J. Chiu, T.~Tzeng, and T.~Lay,
  \enquote{{InGaAs} self-assembly quantum dot for high-speed 1300 nm
  electroabsorption modulator,} {\protect\JournalTitle{Journal of Crystal
  Growth}} \textbf{323}, 473–476 (2011). Proceedings of the 16th
  International Conference on Molecular Beam Epitaxy (ICMBE).

\bibitem{Sobhani}
S.~A. {Sobhani}, B.~J. {Stevens}, N.~{Babazadeh}, K.~{Takemasa}, K.~{Nishi},
  M.~{Sugawara}, R.~A. {Hogg}, and D.~T.~D. {Childs}, \enquote{Proposal for
  common active 1.3 μm quantum dot electroabsorption modulated {DFB} laser,}
  {\protect\JournalTitle{IEEE Photonics Technology Letters}} \textbf{31},
  419–422 (2019).

\bibitem{Luo_2006}
Y.~Luo, B.~Xiong, J.~Wang, P.~Cai, and C.~Sun, \enquote{40 {GHz} {AlGaInAs}
  multiple-quantum-well integrated electroabsorption modulator/distributed
  feedback laser based on identical epitaxial layer scheme,}
  {\protect\JournalTitle{Japanese Journal of Applied Physics}} \textbf{45},
  L1071–L1073 (2006).

\bibitem{doi:10.1063/1.3119186}
C.~Y. Ngo, S.~F. Yoon, W.~K. Loke, Q.~Cao, D.~R. Lim, V.~Wong, Y.~K. Sim, and
  S.~J. Chua, \enquote{Characteristics of 1.3 μm {InAs/InGaAs/GaAs} quantum
  dot electroabsorption modulator,} {\protect\JournalTitle{Applied Physics
  Letters}} \textbf{94}, 143108 (2009).

\bibitem{1512276}
M.~Areiza, C.-B. Tribuzy, S.~Landi, M.~Pires, and P.~Souza, \enquote{Amplitude
  modulators containing an nipi delta doping superlattice,}
  {\protect\JournalTitle{IEEE Photonics Technology Letters}} \textbf{17},
  2071–2073 (2005).

\bibitem{Sandall2013EvaluationOI}
I.~Sandall, J.~Ng, J.~David, H.~Liu, and C.~Tan, \enquote{Evaluation of {InAs}
  quantum dots on {Si} as optical modulator,}
  {\protect\JournalTitle{Semiconductor Science and Technology}} \textbf{28},
  094002 (2013).

\bibitem{Chu:07}
Y.~Chu, M.~Thompson, R.~Penty, I.~White, and A.~Kovsh, \enquote{1.3 μm
  quantum-dot electro-absorption modulator,} in \emph{Conference on Lasers and
  Electro-Optics/Quantum Electronics and Laser Science Conference and Photonic
  Applications Systems Technologies,}  (Optical Society of America, 2007), p.
  CMP4.

\bibitem{Tang:12}
Y.~Tang, J.~D. Peters, and J.~E. Bowers, \enquote{Over 67 {GHz} bandwidth
  hybrid silicon electroabsorption modulator with asymmetric segmented
  electrode for 1.3 μm transmission,} {\protect\JournalTitle{Opt. Express}}
  \textbf{20}, 11529–11535 (2012).

\bibitem{6324086}
P.~Chaisakul, M.-S. Rouifed, D.~Marris-Morini, G.~Isella, D.~Chrastina,
  J.~Frigerio, X.~{Le Roux}, S.~Edmond, J.-R. Coudevylle, and L.~Vivien,
  \enquote{High speed electro-absorption modulator based on quantum-confined
  {Stark} effect from {Ge/SiGe} multiple quantum wells,} in \emph{The 9th
  International Conference on Group IV Photonics (GFP),}  (2012), p. 60–62.

\bibitem{Edwards:13}
E.~H. Edwards, L.~Lever, E.~T. Fei, T.~I. Kamins, Z.~Ikonic, J.~S. Harris,
  R.~W. Kelsall, and D.~A.~B. Miller, \enquote{Low-voltage broad-band
  electroabsorption from thin {Ge/SiGe} quantum wells epitaxially grown on
  silicon,} {\protect\JournalTitle{Opt. Express}} \textbf{21}, 867–876
  (2013).

\bibitem{doi:10.1063/1.118774}
D.~T. Neilson, L.~C. Wilkinson, D.~J. Goodwill, A.~C. Walker, B.~Vögele,
  M.~McElhinney, F.~Pottier, and C.~R. Stanley, \enquote{Effects of lattice
  mismatch due to partially relaxed buffer layers in {InGaAs/AlGaAs} strain
  balanced quantum well modulators,} {\protect\JournalTitle{Applied Physics
  Letters}} \textbf{70}, 2031–2033 (1997).

\bibitem{Millerenergy}
D.~A.~B. Miller, \enquote{Energy consumption in optical modulators for
  interconnects,} {\protect\JournalTitle{Opt. Express}} \textbf{20},
  A293–A308 (2012).

\bibitem{Zhang2018}
Z.~Zhang, D.~Jung, J.~C. Norman, P.~Patel, W.~W. Chow, and J.~E. Bowers,
  \enquote{Effects of modulation \emph{p} doping in {InAs} quantum dot lasers
  on silicon,} {\protect\JournalTitle{Applied Physics Letters}} \textbf{113},
  061105 (2018).

\bibitem{Feng2011}
N.-N. Feng, D.~Feng, S.~Liao, X.~Wang, P.~Dong, H.~Liang, C.-C. Kung, W.~Qian,
  J.~Fong, R.~Shafiiha, Y.~Luo, J.~Cunningham, A.~V. Krishnamoorthy, and
  M.~Asghari, \enquote{30 {GHz} {Ge} electro-absorption modulator integrated
  with 3 μm silicon-on-insulator waveguide,} {\protect\JournalTitle{Opt.
  Express}} \textbf{19}, 7062--7067 (2011).

\bibitem{Srinivasan2020}
S.~A. Srinivasan, C.~Porret, E.~Vissers, P.~Favia, J.~De~Coster, H.~Bender,
  R.~Loo, D.~Van~Thourhout, J.~Van~Campenhout, and M.~Pantouvaki, \enquote{High
  absorption contrast quantum confined {Stark} effect in ultra-thin {Ge/SiGe}
  quantum well stacks grown on {Si},} {\protect\JournalTitle{IEEE Journal of
  Quantum Electronics}} \textbf{56}, 1--7 (2020).

\bibitem{Mahoney:21}
J.~Mahoney, P.~M. Smowton, B.~Maglio, L.~Jarvis, C.~Allford, S.~Shutts,
  M.~Tang, H.~Liu, and N.~Abadía, \enquote{{QCSE} and carrier blocking in
  p-modulation doped {InAs/InGaAs} quantum dots,} in \emph{Conference on Lasers
  and Electro-Optics,}  (Optical Society of America, 2021), p. JTu3A.167.

\bibitem{Cornet2004}
C.~Cornet, C.~Labbé, H.~Folliot, N.~Bertru, O.~Dehaese, J.~Even, A.~{Le
  Corre}, C.~Paranthoen, C.~Platz, and S.~Loualiche, \enquote{Quantitative
  investigations of optical absorption in {InAs∕InP(311)B} quantum dots
  emitting at 1.55 μm wavelength,} {\protect\JournalTitle{Applied Physics
  Letters}} \textbf{85}, 5685–5687 (2004).

\bibitem{PhysRevLett.84.733}
P.~W. Fry, I.~E. Itskevich, D.~J. Mowbray, M.~S. Skolnick, J.~J. Finley, J.~A.
  Barker, E.~P. O'Reilly, L.~R. Wilson, I.~A. Larkin, P.~A. Maksym,
  M.~Hopkinson, M.~Al-Khafaji, J.~P.~R. David, A.~G. Cullis, G.~Hill, and J.~C.
  Clark, \enquote{Inverted electron-hole alignment in {InAs-GaAs}
  self-assembled quantum dots,} {\protect\JournalTitle{Phys. Rev. Lett.}}
  \textbf{84}, 733–736 (2000).

\bibitem{NN}
S.~Birner, \enquote{nextnano: General purpose {3-D} simulations,}
  {\protect\JournalTitle{IEEE Transactions on Electron Devices}} \textbf{54},
  2137–2142 (2007).

\end{thebibliography}

\end{document}